\documentstyle[sprocl]{article}

\def\Journal#1#2#3#4{{#1} {\bf #2}, #3 (#4)}

\def\gd{g_{\mu\nu}}

\def\be{{\beta}}
\def\gam{{\gamma}}

\def\ua{^{\alpha}}

\def\umu{^{\mu}}
\def\unu{^{\nu}}

\def\dmunu{_{\mu\nu}}

\def\ie{{\it i.e. }}

\def\NPB{{\em Nucl. Phys.} B}
\def\PRL{\em Phys. Rev. Lett.}

\def\CQG{\em Class. Quantum Grav.}

\def\PR{\em Phys. Rep.}

\def\be{\begin{equation}}
\def\ee{\end{equation}}
\def\bea{\begin{eqnarray}}
\def\eea{\end{eqnarray}}
\begin{document}

\title{SCALAR PERTURBATIONS IN DEFLATIONARY COSMOLOGICAL MODELS}

\author{S. CAPOZZIELLO, G. LAMBIASE, G. SCARPETTA}

\address{Dipartimento di Scienze Fisiche ``E.R. Caianiello'',
Universit\`a di Salerno,\\I-84081 Baronissi, 
Salerno, ITALY\\E-MAIL: capozziello,lambiase,scarpetta@vaxsa.csied.unisa.it}

\maketitle\abstracts{We consider  scalar perturbations of 
energy--density for a class 
of cosmological models where an early phase of accelerated expansion 
evolves, without
any fine--tuning for graceful exit, towards the standard Friedman eras
of observed universe. The geometric procedure which generates such models
agrees with results for string cosmology since it works if
dynamics is dominated by a primordial  fluid of extended massive objects.
The main result is that characteristic scales of cosmological interest, 
connected with the extension of such early objects, are  selected.}

\noindent Any inflationary model has to satisfy the issue that, during 
the expansion, perturbations which are inside the Hubble radius $H^{-1}$,
at the beginning of inflation, expand, past the Hubble radius, and reenter 
it at
late times as large scale density perturbations.
To calculate the amplitude of density perturbations and to study the 
transition from inflationary to the Friedman era, it is necessary to know 
how the background geometry change with time.
Furthermore, a coherent theory of early universe should
$i)$ be connected to some unification scheme of all interactions of nature;
$ii)$ avoid the initial singularity;
$iii)$ evolve smoothly, \ie without fine--tuning, from an inflationary stage
to a decelerating Friedman era;
$iv)$ give rise to a perturbation spectrum in agreement both with the observed
microwave background isotropy and with the large scale structures.
In other words,  a cosmological model,
 connected with some fundamental theory, should, at a certain epoch,
acquire a deflationary behaviour ~\cite{barrow} and reproduce a suitable
perturbation spectrum.
Using a quantum geometric procedure 
~\cite{caianiello1}, we construct a class of cosmological models of 
deflationary type wich smoothly evolves towards Friedman epochs.
Over this background, it is possible to analyze the theory of gauge 
invariant cosmological 
perturbations for the density contrast 
$\delta \rho/\rho$ connecting it with the scales of
astrophysical interest.

The main hypothesis to build such models is that the early universe is
dominated by a fluid of finite--size objects which give rise to a
dynamics very similar to that of string--dilaton cosmology 
~\cite{tseytlin}. However, the starting point is different from
that of string theory since our procedure is just a quantum 
geometric scheme.
We do not need any scalar field to implement inflation, since the
proper size of extended objects, and the geodesic embedding procedure from an 
eight--dimensional tangent fiber bundle ${\cal M}_{8}$ to the usual
${\cal V}_{4}$ manifold of general relativity naturally give rise to an 
exponential inflationary--like behaviour.
The quantum geometric procedure and 
the background model which we are going to use
are treated in detail in ~\cite{caianiello1}. Here, we outline the main 
features which we need for cosmological perturbations ~\cite{pertacc}.
The starting point
is that if we consider dynamics of an extended massive object in general 
relativity, a limiting maximal acceleration, compatible with the size
$\lambda$ of the object and the causal structure of the spacetime manifold, 
emerges ~\cite{misner}.
Such a proper constant acceleration $A$ yields a Rindler horizon at a 
distance $|A|^{-1}$ from the extremity of the object in the longitudinal
direction. In other words, the parts of the object will be in causal contact
only if $|A|<\lambda^{-1}$. It is worthwhile to stress that the parameter
$A$ (or $\lambda$) is related to the "mass" of the extended object. 
Let us take into account a Friedman--Robertson--Walker (FRW) spacetime
whose scale factor, with respect to the cosmic time $t$, is $a(t)$.
By using the equation of geodesic deviation 
~\cite{caianiello1}, we get that 
the size $\lambda$ of the object is compatible with the causal structure if
$|\lambda \ddot{a}/a|<1$.
Consequently, we have a maximal allowed curvature depending on $\lambda$
and the cosmological model becomes singularity free.
This fact is in sharp contrast with usual perfect fluid FRW cosmology where 
curvature and matter--energy density are singular in the limit $t\rightarrow 0$.
Then, the introduction of finite size objects, instead of pointlike 
particles, in primordial cosmological background, modifies dynamics so that 
the singular structure of general relativity is easily regularized.
It is worthwhile to note that such a feature does not depend on the 
particular background geometry which we are considering.
More formally, a causal structure in which proper accelerations cannot exceed
a given value $\lambda^{-1}$ can be imposed over a generic spacetime
${\cal V}_{4}$ regarding such a manifold as a four--dimensional
hypersurface locally embedded into an eight--dimensional tangent fiber bundle
${\cal M}_{8}$, with metric
$g_{AB}=\gd\otimes\gd,$
and coordinates $x^{A}=(x\umu,\lambda u\umu)$, where 
$ u\umu=dx\umu/ds$ is the usual four velocity
and $\mu,\nu=1,...,4,$ $A,B=1,...,8$ 
~\cite{caianiello2}.
The embedding of ${\cal V}_{4}$ into ${\cal M}_{8}$, determined by the eight
parametric equations $x\umu=x\umu(\xi\ua)$ and $u\umu=u\umu(\xi\ua)$,
gives rise to a spacetime metric $\tilde{g}\dmunu(\xi)$, locally induced by 
the ${\cal M}_{8}$ invariant interval
\be
\label{2.1}
d\tilde{s}^2=g_{AB}dx^Adx^B=
\gd\left(dx\umu dx\unu+\lambda^2du\umu du\unu\right)
\equiv\tilde{g}\dmunu d\xi\umu d\xi\unu\,.
\ee
Let us now take into consideration a FRW flat--background modified by such a 
geodesic embedding.
In conformal coordinates $\xi\umu=(\eta,\vec{x})$, we get ~\cite{caianiello1}
\be
\label{2.5}
\tilde{g}\dmunu(\xi)=
\mbox{diag} a^2\left(1+\lambda^2\frac{{a'}^2}{a^4},-1,-1,-1\right)\,.
\ee
The prime indicates the derivative with respect to $\eta$.
The cosmic time results now
\be
\label{2.6}
t=\lambda\int\frac{da}{a}\left(1+\frac{a^4}{\lambda^2{a'}^{2}}\right)^{1/2}\,.
\ee
The Hubble parameter is 
\be
\label{2.8}
H=\frac{\dot{a}}{a}=\left(\frac{a'}{a}\right)
\left[a^2+\lambda^2\left(\frac{a'}{a}\right)^2\right]^{-1/2}\,,
\ee
with the limiting value $H_{0}=\lambda^{-1}$
for $\lambda^{2}(a'/a)^2\gg a^2$.
It is easy to see that the scale factor, with respect to the cosmic time
$t$, has an initial exponential growth
which regularly evolves towards a standard Friedman behaviour.
The $e$-folding number,
\ie the duration of inflation, and the horizon scale depend on the size 
$\lambda$ without any initial value problem or fine tuning.
The natural scale to which to compare perturbations is $\lambda$:
they are inside the Hubble radius if they are smaller than $\lambda$
while they are outside it if they are greater than $\lambda$.
In other words, $\lambda$ determines the crossing time
(either out of the Hubble radius or into the Hubble radius).
Considering the $(0,0)$--Einstein equation for a spatially flat model,
we have 
$H^2=\rho/3$
so that
\be
\label{3.3}
\rho\simeq\frac{3}{\lambda^2}\,,\;\;\;\mbox{for}\;\;\;\;
\lambda^2\left(\frac{a'}{a}\right)^{2}\gg a^2\,;\;\;\;\;
\mbox{and}\;\;\;\;
\rho\simeq 3\frac{{a'}^{2}}{a^4}\,,\;\;\;\;\mbox{for}\;\;\;
\lambda^2\left(\frac{a'}{a}\right)^2\ll a^2\,.
\ee
The first case corresponds to an effective cosmological constant
$\Lambda=3/\lambda^2$ selected by the mass 
(\ie the size) of the primordial extended objects ~\cite{caianiello1}; 
the second case
is recovered as soon as 
the universe undergoes the post--inflationary reenter phase. 
We stress again the fact that such
a behaviour does not depend on the specific form of the scale factor $a$
and the deflationary phase is smooth.
As in ~\cite{caianiello1}, we can couple dynamics with ordinary fluid matter in
order to obtain a more realistic cosmological scenario. In doing so, 
we have to consider a perfect fluid state equation
$p=(\gamma-1)\rho$
which, using also the contracted Bianchi identity in FRW spacetime gives
the continuity equation
$\dot{\rho}+3H(p+\rho)=0\,,$
from which we get
$\rho=Da^{-3\gamma}.$
As usual, $\gamma$ is assumed constant. It defines
the thermodynamical state of the fluid and it is related to the sound speed
being $\gam-1=c_{s}^{2}$. 
By inserting this fluid into the Einstein equations, 
the scale factor of the universe, expressed in conformal time is
~\cite{turok}
\be
\label{solu}
a(\eta)=a_{0}\eta^{2/(3\gamma-2)}\,,
\ee
where $a_{0}$ is a constant depending on $\lambda$ and $\gamma$.
The matter--energy density results, 
\be
\label{3.10}
\rho=3\left(\frac{2}{3\gamma -2}\right)^2\frac{1}{\eta^2}
 \left[a_{0}^2\eta^{\frac{4}{3\gamma-2}}+
\lambda^{2}\left(\frac{2}{3\gamma-2}\right)^2\frac{1}{\eta^2}\right]^{-1}\,,
\ee
from which $\rho\sim$ constant for 
$\lambda\eta^{6\gamma/(2-3\gamma )}\gg 1$ and 
$\rho\sim\eta^{\frac{6\gamma}{2-3\gamma}}$ in the opposite
case. 
The standard situations for $\gamma=4/3$ (radiation dominated regime)
and $\gamma=1$ (matter dominated regime) are easily recovered.
It is interesting to see that it is not only the specific value of 
$\gamma=0$, as usual, that allows to recover inflation but, mainly,
the scale $\lambda$. In the regime 
$\lambda/\eta\gg 1$, the constant matter density
value is independent of $\gamma$.
The general gauge--invariant evolution equation for 
scalar adiabatic perturbations
is ~\cite{mukhanov}
\be
\label{master}
\Phi''+3{\cal H}(1+c_{s}^{2})\Phi'
-c_{s}^{2}\nabla^{2}\Phi+
\left[2{\cal H}'+(1+3c_{s}^{2}){\cal H}^{2}\right]\Phi=0\,,
\ee
where $\Phi$ is the perturbation potential,
${\cal H}$ is the Hubble parameter in the conformal time defined as
${\cal H}=a'/a $. It is interesting to note that Eq.(\ref{master}) can 
be recast in terms of
the scale factor $a$ by the variable change $dt=a\,d\eta$ and $da/dt=aH$
~\cite{caldwell}. In this way, the information contained in the evolution
equation is directly related to the background. However, for our purposes,
it is better to use the "conformal time picture" 
since it immediately shows when the  
sizes of perturbations are comparable to the characteristic scale 
length $\lambda$.
For example, if $k\ll H$, we have long wavelength modes which furnish the 
spectrum of perturbations during inflation. In our case, it is interesting
to compare such modes with the "natural" scale of the model, that is
$H_{0}=\lambda^{-1}$.
Before performing the Fourier analysis, it is useful to simplify the
dynamical problem
by a suitable change of variables. Eq.(\ref{master}) 
can be reduced to the simpler form
\be
\label{master1}
u'' - c^2_s\nabla^2 u - {\theta''\over\theta}u=0\,,
\ee
where $\theta$ is
\be
\theta={1\over a}\left(\frac{\rho_0}{\rho_0 + p_0}
\right)^{1/2}={1\over a}\left(\frac{1}{1 +  
p_0 /\rho_{0}}\right)^{1/2}=\frac{1}{a\sqrt{\gamma}}\,,
\ee
and the gauge--invariant gravitational potential $\Phi$ is given  by 
\be
\Phi= {1\over2}(\rho_0+ p_0)^{1/2}u\,.
\ee
From now on, the subscript $"_0"$ will indicate the unperturbed quantities.
In the specific case  we are considering, using the solution (\ref{solu}),
we get
\be
\label{theta}
\theta(\eta)=\left[\frac{2}{H_{0}\gamma^{1/2}(3\gamma-2)}\right]
\eta^{2/(3\gamma-2)}\,.
\ee
After the Fourier trasform, Eq.(\ref{master1}) becomes
\be
\label{master2}
u''_{k}+
\left[c_{s}^2 k^2-\frac{6\gamma}{(3\gamma-2)^{2}\eta^2}\right]u_{k}=0\,,
\ee
which is nothing else but a Bessel equation.
The density perturbations can be written as
\be
\label{contrast1}
\frac{\delta\rho}{\rho_{0}}=-(\rho_{0}+p_{0})^{1/2}
\left[\left(1+\frac{k^2}{{\cal H}^2}\right)u_{k}(\eta)+
\frac{u'_{k}(\eta)}{\cal H}\right]\,,
\ee 
where
\be
\label{hubble}
{\cal H}=\left(\frac{2}{3\gamma-2}\right)\frac{1}{\eta}\,.
\ee
The general solution of (\ref{master2}) is
\be
\label{solution}
u_{k}(\eta)=\eta^{1/2}\left[A_{0}J_{\nu}(z)+B_{0}Y_{\nu}(z)\right]\,,
\ee
where $J_{\nu}(z)$ and $Y_{\nu}(z)$ are Bessel functions and
\be
\nu=\pm\frac{3\gamma+2}{2(3\gamma-2)}\,,\;\;\;\;\;\;\;
z=c_{s}k\eta\,; 
\ee
$A_{0}$ and $B_{0}$ are constants. 
The large scale limit is recovered as soon as $k^2\ll 
\theta''/\theta$, or $k\ll H$.
This means that the solution (\ref{solution}) becomes
\be
\label{lsl}
u_{k}(\eta)\simeq \eta^{1/2}\left[\frac{A_{0}}{\Gamma(\nu+1)}
\left(\frac{c_{s}k\eta}{2}\right)^{\nu}-
\frac{B_{0}\Gamma(\nu)}{\pi}\left(\frac{c_{s}k\eta}{2}\right)^{-\nu}
\right]\,.
\ee
For different values of $\gamma$, the index $\nu$ can be positive or negative
determininig growing or decaying modes. 
In the vacuum--dominated era $(\gamma=0)$, we have, for $k\rightarrow 0$,
\be
u_{k}(\eta)\sim\left[ B_{0}\sqrt{\frac{2}{\pi c_{s}}}\right]k^{-1/2}\,,
\;\;\;\;\;
\mbox{or}
\;\;\;\;\;
\frac{\delta\rho}{\rho_{0}}\sim 
\left[B_{0}\sqrt{\frac{2(\rho_{0}+p_{0})}{\pi c_{s}}}\right]k^{-1/2}\,.
\ee
This is a nice feature since the spectrum of perturbations is a constant
with respect to $\eta$ as it must be during inflation, when dynamics
is frozen ~\cite{kolb}. As we pointed out, we recover the case $\gamma=0$
any time that $H_{0}=\lambda^{-1}=k_{\lambda}$, that is the feature of the 
spectrum is fixed by the natural scale of the model
If $\gamma$ is any, in particular $\gamma=1,4/3,2$, 
corresponding to the 
cases "dust", "radiation" 
and "stiff matter" 
respectively, we get
\be
\label{matter}
u_{k}(\eta)\sim \frac{A_{0}}{\Gamma(\nu+1)}\left(\frac{c_{s}}{2}\right)^{\nu}
\eta^{(\nu+1/2)}k^{\nu}-\frac{B_{0}\Gamma(\nu)}{\pi}
\left(\frac{c_{s}}{2}\right)^{-\nu}\eta^{(1/2-\nu)}k^{-\nu}\,.
\ee
In particular, for $k\rightarrow 0$, only the second term survives.
The density contrast, in the same limit, is
\be
\label{contrast3}
\frac{\delta \rho}{\rho_{0}}\sim\frac{B_{0}\Gamma(\nu)}{\pi}
\sqrt{\rho_{0}+p_{0}}\left(\frac{2}{c_{s}}\right)^{\nu}
\frac{(3\gamma-2)^2}{12}\eta^{(\frac{5}{2}-\nu)}k^{(2-\nu)}\,.
\ee
It is interesting to note that, for $\gamma=1$, $\nu=5/2$ and $
{\delta\rho}/{\rho_{0}}\propto k^{-1/2},$
that is we lose the time dependence also if the scales are reentered the
horizon (for $\gamma=1$ we are in the Friedman regime).
The small scale limit is recovered as soon as 
$k^{2}\gg \theta''/\theta$. The solution can be written as
\be
u_{k}(\eta)\sim\sqrt{\frac{2}{\pi c_{s}k}}\left[A_{0}\cos(c_{s}k\eta)+
B_{0}\sin(c_{s}k\eta)\right]\,,
\ee
so that the density contrast is an 
 oscillating function in $\eta$.
In conclusion, the main point of our model is that the size of the 
objects gives rise to
an inflationary period that smoothly evolves toward a Friedman era.
As a consequence, the cosmological perturbations are affected 
by such a dynamics since
the scales are regulated by the size $\lambda$
which gives the Hubble horizon $H_{0}=\lambda^{-1}$
during inflation.
Then the limits to compare {\it very} large scale structures and
{\it small} large scale structures are $k\ll H_{0}$ and $k\gg H_{0}$.
In other words, $\lambda$ fixes the time at which perturbations
cross the horizon and reenter into it without any
fine--tuning.

\end{document}